\documentclass{article}

\usepackage{arxiv}

\usepackage[utf8]{inputenc} 
\usepackage[T1]{fontenc}    
\usepackage{hyperref}       
\usepackage{url}            
\usepackage{booktabs}       
\usepackage{amsfonts}       
\usepackage{nicefrac}       
\usepackage{microtype}      
\usepackage{graphicx}
\usepackage{epstopdf, epsfig}
\usepackage{subfigure}
\usepackage{amsmath,amssymb}

\title{A two layer model for wave dissipation in sea ice}
\author{
  Graig Sutherland\\
  Numerical Environmental Prediction Research\\
  Environment and Climate Change Canada\\
  \texttt{graigory.sutherland@canada.ca} \\
   \And
  Jean Rabault \\
  Department of Mathematics\\
  University of Oslo\\
  \texttt{jeanra@math.uio.no} \\
   \And
  Kai H. Christensen \\
  Norwegian Meteorological Institute\\
  \texttt{kaihc@met.no} \\
   \And
  Atle Jensen \\
  Department of Mathematics\\
  University of Oslo\\
  \texttt{atlej@math.uio.no} \\
}

\begin{document}
\maketitle

\begin{abstract}


   Sea ice is highly complex due to the inhomogeneity of the physical properties (e.g. temperature and salinity) as well as the permeability and mixture of water and a matrix of sea ice and/or sea ice crystals. Such complexity has proven itself to be difficult to parameterize in operational wave models. Instead, we assume that there exists a self-similarity scaling law which captures the first order properties. Using dimensional analysis, an equation for the kinematic viscosity is derived, which is proportional to the wave frequency and the ice thickness squared. In addition, the model allows for a two-layer structure where the oscillating pressure gradient due to wave propagation only exists in a fraction of the total ice thickness. These two assumptions lead to a spatial dissipation rate that is a function of ice thickness and wavenumber. The derived dissipation rate compares favourably with available field and laboratory observations.

\end{abstract}




%

\section{Introduction}

The marginal ice zone (MIZ) can be defined as the interfacial region of the cryosphere between the open ocean and consolidated pack ice, and is the region of the cryosphere that is most affected by surface ocean waves. The MIZ is a complex and heterogeneous environment, which comprises of ice floes of arbitrary shape and a variety of ice types from newly formed ice to fragmented multi-year ice. While surface waves are expected to be a dominant process controlling the total extent of the MIZ as well as the size distribution of ice floes, there still exist large uncertainties about the exact mechanisms for wave attenuation in sea ice~\citep{Williams_etal_2013a,Williams_etal_2017,Kohout_etal_2014,Meylan_etal_2014}. In addition to affecting the floe size distribution, the attenuation of surface gravity waves is also important in controlling the extent of the MIZ via the wave radiation stress~\citep{Liu_etal_1993,Williams_etal_2017,PSutherland_Dumont_2018}.

The two predominant sources for the attenuation of wave energy propagating into an ice field are due to scattering, which is a conservative process that redistributes energy in all directions, and dissipative processes due to friction (viscosity), inelastic collisions and the breakup of ice floes~\citep{Squire_1995, Squire_2007,Doble_etal_2013,Williams_etal_2013a}. The relative importance of scattering and dissipative processes is still unclear and a topic of active research~\citep{Squire_Montiel_2016}. Theoretical studies of wave scattering suggest that dissipative mechanisms may be dominant for small floe sizes~\citep{Kohout_Meylan_2008,Bennetts_Squire_2012}, relative to the incoming wavelength, as would be found in frazil and pancake ice fields in the MIZ~\citep{Doble_etal_2015}. In addition, \citet{Ardhuin_etal_2016} found that dissipative mechanisms dominated the wave attenuation for narrow banded swell travelling several hundred kilometres into the pack ice~\citep{Wadhams_Doble_2009}, which suggests that wave scattering by ice floes is more complex. Recent studies on wave attenuation in the MIZ have focused on dissipative processes~\citep{Doble_etal_2015,Rogers_etal_2016,Cheng_etal_2017} as these are expected to be dominant in the MIZ.

There are many factors which contribute to the difficulty of determining wave attenuation in sea ice. First, field observations are difficult to obtain and thus remain relatively sparse. To obtain estimates of the wave attenuation requires collocated observations of the wave amplitude and wave direction at several frequencies over spatial distances on the order of kilometres. In addition, it is not only the sea ice which affects the spectral energy of the surfaces waves as the wind forcing and the nonlinear transfer of energy can also represent significant contributions in the lower ice concentrations found in the MIZ~\citep{Li_etal_2015,Li_etal_2017}. Second, while laboratory experiments of waves and sea ice circumvent many of the problems encountered in interpreting field data, there exist many problems associated with scaling laboratory experiments to larger scales. For instance, it is desirable to not have the bottom influence surface experiments, which requires $kD>1$ where $k$ is the wavenumber and $D$ is the water depth. Furthermore, it is also desirable to perform experiments with relatively linear waves and thus limiting the wave steepness to $ak<0.1$ where $a$ is the wave amplitude. For a wave tank of depth $D=1$ m, these two criteria require amplitudes to be less than 0.1 m and frequencies to be greater than 0.5 Hz. This is why most laboratory experiments are performed with low amplitude and high frequency waves. See~\citet{Rabault_etal_2018_jfm} for a discussion on difficulties in scaling laboratory experiments.

There exist many models that describe viscous dissipation in a frazil and pancake ice field, with various degrees of consistency with observations. \citet{Weber_1987} assumed that the ice layer could be modelled as a highly viscous continuum where the dynamics in the ice layer consist of a balance between pressure and viscous forces, i.e. ``creeping motion'' or Stokes flow (not to be confused with ice creep~\citep[e.g.][]{Wadhams_1973}), and the wave dissipation exists solely in the ocean boundary layer below the sea ice. In this case, \citet{Weber_1987} showed that the sea ice layer effectively halts the horizontal motion of the water at the ice-water interface and the spatial dissipation is identical to the inextensible (i.e. non-elastic) limit for waves under a surface cover~\citep{Lamb_1932,Phillips_1977,Sutherland_etal_2017}. This parameterization compared well with available field observations in the MIZ at the time~\citep{Wadhams_etal_1988} with an eddy viscosity under the ice between 2 and 4 orders of magnitude greater than the molecular value for sea water. While this model has been shown to also compare well with later observations~\citep{Sutherland_Gascard_2016,Rabault_etal_2017}, it does require the fitting of an eddy viscosity, which is required to vary by several orders of magnitude to be consistent with observations and such large values may not be physically realistic~\citep{Stopa_etal_2016}. More recently, \citet{Marchenko_Chumakov_2017} and~\citet{Marchenko_2018} extended the theory of~\citet{Weber_1987} to account for a partial-slip boundary condition and to include floe-floe interactions. The eddy viscosity, and partial-slip parameter, are tuned to observations in the MIZ, and good agreement is found with the assumption of no-slip, or near no-slip, and an eddy viscosity 4 orders of magnitude greater than the molecular value.

Other studies on wave dissipation have focused on the dissipation of wave energy within the sea ice, and modelled the sea ice as a viscous~\citep{Keller_1998,DeCarolis_Desiderio_2002} or viscoelastic~\citep{Wang_Shen_2010} layer over an inviscid~\citep{Keller_1998,Wang_Shen_2010} or viscid ocean~\citep{DeCarolis_Desiderio_2002}. While there exist consistencies between these models and available observations~\citep[e.g.][]{Newyear_Martin_1999}, all of these models require the rheological properties of the sea ice to be empirically determined from observations of wave dissipation. In addition, these models require solving a complex dispersion relation to determine the most physically significant solution, which is not always straightforward and simple~\citep[see][]{Mosig_etal_2015}. The complexity of these models increase the computational expense of implementation into numerical prediction systems, in addition to having a tendency of obscuring the physical interpretation. 

There has been significant work recently on using available observations for testing~\citep{Rogers_etal_2016} and calibrating~\citep{Cheng_etal_2017} the viscoelastic model of~\citet{Wang_Shen_2010} to be implemented in WAVEWATCH III\textsuperscript{\textregistered}~\citep{WW3_v516}. The viscoelastic model is particularly appealing, in theory, due to the use of one rheological model to predict wave propagation in all types of sea ice. This ultimately requires careful tuning of the viscous and elastic parameters over a large variety of sea ice types and conditions. However, field~\citep{Rogers_etal_2016,Cheng_etal_2017} and laboratory~\citep{Wang_Shen_2010_exp,Zhao_Shen_2015} studies have struggled to predict wave attenuation across a broad range of frequencies and ice types. For frazil and pancake ice, these studies also require a non-zero elastic term in order to be consistent with observations, but as of yet there does not exist a strong physical explanation for the inclusion of such an elastic term, which must be tuned from available observations of wave attenuation. Due to the requirement of empirically fitting the sea ice rheology, the viscoelastic model also provides an extra term for fitting, and thus should provide improved consistency with observations independent of whether the physics are correct. 


The theories described here all assume the sea ice has a vertically homogeneous structure, which is rare in reality as there exists large vertical gradients in temperature, and often the bulk salinity, within the sea ice~\citep{Cox_Weeks_1974}. The temperature and bulk salinity, along with the related fraction of solid ice parameter, affect many properties of the sea ice~\citep{Hunke_etal_2011}, for example porosity and permeability, that it seems natural that this vertical structure will also affect wave propagation. In this paper, a new parameteriztion for wave dissipation is presented, which allows for a two-layer structure within the ice. This model assumes wave motion, i.e. an oscillating pressure gradient, in a fraction of the sea ice and that the remainder of the sea ice is too viscous to render it effectively solid over the temporal scales of typical ocean waves. The viscosity is derived using dimensional analysis, which assumes the existence of a universal self-similar scaling to explain the complex phenomenon associated with wave propagation in different ice types. The outline of the paper is as follows. Theoretical details of the two layer wave in ice model is given in Section~\ref{sec:theory}. This is followed by comparisons with available laboratory and field experiments in section~\ref{sec:results}. A discussion of the results can be found in section~\ref{sec:discussion}, along with a summary of the main conclusions in section~\ref{sec:conclusions}.

\section{Wave dissipation in sea ice}
\label{sec:theory}

\begin{figure}[t]
   \centering
   \includegraphics[angle=0]{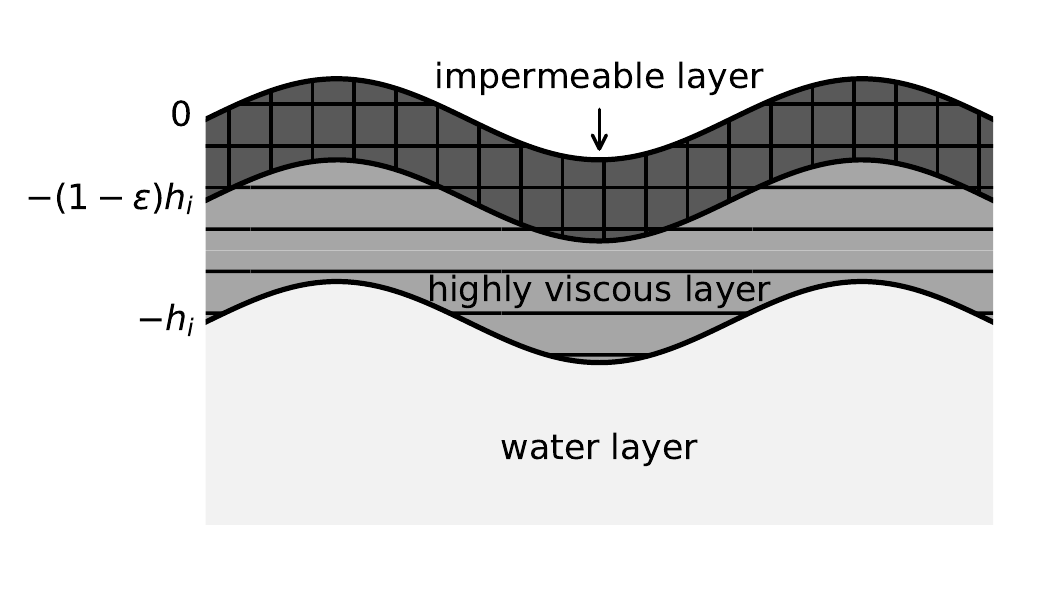} 
\caption{Schematic of the two-layer sea ice model.}
\label{fig:schematic}
\end{figure}

The sea ice is assumed to be locally horizontally homogeneous, i.e. horizontal variability occurs on scales much greater than the wavelength, and is permitted to have a two layer structure, as shown in Figure~\ref{fig:schematic}. The two layer model assumes there is a lower layer where wave motion exists, i.e. where an oscillating pressure gradient exists, and an upper layer which does not permit such motion. We define the thickness of the wave permitting lower layer to be $\epsilon h_i$, where $h_i$ is the total thickness of the sea ice. This implies that this theory is readily compatible with previous theories by setting $\epsilon \to 0$ to have no wave-permitting region, such as in~\citet{Weber_1987}, and $\epsilon \to 1$ to have all wave-permitting such as in~\citet{Keller_1998}. We also assume the periodic wave motion is in the $x-z$ plane, where $x$ is the horizontal direction and $z$ is the vertical direction. 

It is assumed that the periodic velocity associated with the surface waves, $u$, can be separated into rotational and irrotational components,
\begin{equation}
	u = \frac{\partial \phi}{\partial x} + u^{\prime},
  \label{eq:u}
\end{equation}
where $\phi$ is the velocity potential and $u^{\prime}$ is the rotational component. In (\ref{eq:u}) it is implied that $u$, $\phi$ and $u^{\prime}$ are functions of space and time.  For deep water waves the velocity potential is given by $\phi = a c_p \sin{(kx - \omega t)}$ where $a$ is the amplitude, $c_p = \omega / k$ is the phase velocity, $\omega$ is the angular frequency, $k$ is the wavenumber, and $t$ is time. While this analysis is for a monochromatic wave, it can easily be extended to individual components of a broadband wave spectrum by averaging over a sufficiently long period to eliminate low-frequency oscillations due to wave components closely spaced in frequency~\citep{Chang_1969}. 

All the following analysis will assume that $kh_i \ll 1$ and thus ignore the exponential attenuation of the surface waves with depth. For a 50 m wavelength and 1 m ice thickness the value is $kh_i = 0.1$. Since the most energetic waves in the ocean typically have larger wavelengths, and the wave affected regions will have slightly smaller ice thickness, then our assumption is valid for the vast majority of wave-ice interactions in the ocean.

Below the ice the wave motion is approximately irrotational,
\begin{equation}
	u = \frac{\partial\phi}{\partial x}, \quad z \le -h_i.
\label{eq:bc1}
\end{equation}
In the lower layer, the rotational component is allowed to vary with height,
\begin{equation}
	u(z) = \frac{\partial\phi}{\partial x} + u^\prime(z), \quad -h_i < z < -(1-\epsilon)h_i.
\label{eq:bc2}
\end{equation}
Since $kh_i \ll 1$ we can safely omit the exponential decay with depth of the irrotational component of the wave velocity. In the upper layer, the ice is essentially a deformable solid body and there are no velocity gradients within this region.

Rotational shear only exists in the wave permitting region over the depth range $-h_i < z < -(1-\epsilon)h_i$. Therefore, the dissipation of wave energy can be calculated from the irrotational shear~\citep{Phillips_1977} over this region,
\begin{equation}
\label{eq:energy}
\frac{\partial E}{\partial t} = -\rho\nu \int_{-h_i}^{-(1-\epsilon)h_i}\overline{ \left(\frac{\partial u^\prime}{\partial z}\right)^2}dz,
\end{equation}
where $E$ is the wave energy, $\rho$ is the viscous layer density and $\nu$ is the effective kinematic viscosity associated with the dissipation of wave energy.  To keep the solution general, we assume that velocity in the wave permitting region will have the same structure as the irrotational solution, but with a decrease in amplitude and a phase lag, i.e.
\begin{equation}
   u = \Gamma a\omega\cos{(kx - \omega t - \psi)},\quad -h_i < z < -(1-\epsilon) h_i,
   \label{eq:bc3}
\end{equation}
where $0\le\Gamma \le 1$ is the amplitude reduction and $\psi$ is the phase lag, and are allowed to be functions of $z$. From (\ref{eq:bc3}) we can estimate the rotational component of the velocity
\begin{equation}
	u^\prime = u - \frac{\partial\phi}{\partial x} = \Gamma a\omega\cos{(kx - \omega t - \psi)} - a\omega\cos{(kx - \omega t)}.
   \label{eq:bc3b}
\end{equation}
Assuming that the phase lag $\psi$ is small, such that second order effects can be neglected, (\ref{eq:bc3b}) becomes
\begin{equation} \label{eq:bc3c}
   \begin{split}
      u^\prime & = a\omega \left[ (\Gamma - 1) \cos{(kx - \omega t)} + \Gamma\psi\sin{(kx - \omega t)}\right] \\
											  & = a\omega \left[\sqrt{(\Gamma-1)^2 + (\Gamma\psi)^2}\sin{(kx -\omega t - \theta)}\right],
   \end{split}
\end{equation}
where $\theta = \tan^{-1}(\Gamma - 1) / (\Gamma\psi).$

Now, in order to simplify the problem, we will estimate the mean shear over the wave permitting region of thickness $\epsilon h_i$ using boundary conditions (\ref{eq:bc3c}) and (\ref{eq:bc1}), therefore,
\begin{equation} \label{eq:meanshear}
	\frac{\partial u^\prime}{\partial z} = \frac{aw\left[ (\Gamma - 1) \cos{(kx - \omega t)} + \Gamma\psi\sin{(kx - \omega t)}\right]}{\epsilon h_i}.
\end{equation}
By taking $\Gamma=\Gamma_0$ and $\psi=\psi_0$ evaluated at $z=-(1-\epsilon)h_i$, we can estimate the integral of (\ref{eq:energy}) to be 
\begin{equation}
   \int_{-h_i}^{0}\overline{\left(\frac{\partial u^\prime}{\partial z}\right)^2}dz = \frac{1}{2}\left(\frac{a\omega}{\epsilon h_i}\right)^2 \left[1 - 2\Gamma_0 + (1+\psi_0^2)\Gamma_0^2\right]\epsilon h_i.
   \label{eq:dudz}
\end{equation}

Substituting (\ref{eq:dudz}) into (\ref{eq:energy}), defining $\Delta_0 = 1 - 2\Gamma_0 + (1+\psi_0^2)\Gamma_0^2$, and using the definition for wave energy $E=\rho g a^2/2$, where $g$ is the acceleration due to gravity, allows for (\ref{eq:energy}) to be written as
\begin{equation}
  \label{eq:energy2}
  \frac{\partial E}{\partial t} = -\frac{\nu\omega^2\Delta_0}{g\epsilon h_i}E.
\end{equation}
Using the definition for the temporal dissipation rate, 
\begin{equation}
  \label{eq:beta1}
  \beta = -\frac{1}{2E}\frac{\partial E}{\partial t},
\end{equation}
 along with (\ref{eq:energy2}) and assuming the dispersion relation to be that of open water, $\omega^2=gk$, allows for (\ref{eq:beta1}) to be written as
\begin{equation}
  \label{eq:beta}
  \beta = \frac{\nu k\Delta_0}{2\epsilon h_i}.
\end{equation}
Equation (\ref{eq:beta}) is equivalent to the classical equation for wave dissipation under an inextensible surface cover~\citep{Lamb_1932} assuming a no-slip boundary condition at $z = -(1-\epsilon)h_i$ (i.e. $\Delta_0\rightarrow 1$), and with a boundary layer thickness of $\epsilon h_i$. 

Assuming the wave dissipation is small, such that $\beta \ll \omega$, the temporal dissipation rate can be related to the spatial dissipation rate, which we will denote $\alpha$, using the group velocity, i.e. $c_g = \beta / \alpha$~\citep{Gaster_1962}. Using the group velocity in open water, $c_g = \omega / (2k)$, allows (\ref{eq:beta}) to be converted to a spatial dissipation rate
\begin{equation}
   \label{eq:alpha}
   \alpha = \frac{\nu \Delta_0 k^2}{\omega \epsilon h_i}.
\end{equation}


A large uncertainty in estimating $\alpha$ using (\ref{eq:alpha}) is accurate knowledge of the ice viscosity $\nu$, which is undoubtedly a complicated function of the ice micro-structure. However, some insight might be gained by looking at the relevant macroscopic ice properties with regards to wave dissipation in order to look for a self-similar, i.e. non-dimensional, solution. At the macroscopic level the sea ice can be summarized by two factors: the viscosity $\nu$ and the thickness $h_i$. To first order, and thus ignoring nonlinear effects and using the assumption that $k h_i \ll 1$, the relevant parameters for wave dissipation are then $\nu$, $h_i$ and $\omega$. According to the $\Pi$ theorem~\citep{Buckingham_1914}, there only exists one non-dimensional value and the ice viscosity must scale as
\begin{equation}
   \label{eq:visc0}
   \nu = A \omega h_i^2,
\end{equation}
where $A$ is a constant. 

We will use a constant $A = 0.5 \epsilon^2$ in order to equate (\ref{eq:visc0}) with the classic Stokes boundary layer, $d = \sqrt{2\nu_0/\omega}$ where $\nu_0$ is the (eddy) viscosity of the water~\citep{Lamb_1932,Phillips_1977}, for a boundary layer thickness of $\epsilon h_i$. The ice viscosity is then
\begin{equation}
   \label{eq:visc}
   \nu = \frac{1}{2}\omega\left(\epsilon h_i\right)^2.
\end{equation}
Substituting (\ref{eq:visc}) into (\ref{eq:alpha}) gives
\begin{equation}
  \label{eq:alpha1}
   \alpha = \frac{1}{2}\Delta_0\epsilon h_i k^2.
\end{equation}
Equation (\ref{eq:alpha1}) gives an estimate of the spatial decay rate of surface waves as a function of the wavenumber and ice thickness, with two unknowns, $\Delta_0$ and $\epsilon$, that range in value between zero and one. However, if we are to assume self-similarity for wave dissipation in ice layers, then $\alpha \propto h_i k^2$ should be valid and the constant of proportionality can be determined from experiments. That being said we believe there is some physical interpretation to the constant of proportionality, which is explored next.

\subsection{Determining $\Delta_0$}

The $\Delta_0$ term determines the boundary condition imposed on the bottom of the ice. A no-slip boundary condition, i.e. $\Gamma_0\to 0$, leads to $\Delta_0 = 1$. This was shown by \citet{Weber_1987} to be valid to first order for a highly viscous ice layer, and provided consistent results with observations in the MIZ made by~\citet{Wadhams_etal_1988}. However, \citet{Ardhuin_etal_2018} suggest weakening the no-slip condition depending on the ratio of the floe size to the wavelength and devise an empirical formulation which dramatically reduces the attenuation rate for wavelengths greater than 3 times the dominant floe size. In this formulation there is no explicit mentioning whether this reduction is due to the amplitude, phase lag, or both of the ice motion relative to the water. 

\begin{figure}[t]
   \centering
   \includegraphics[angle=0]{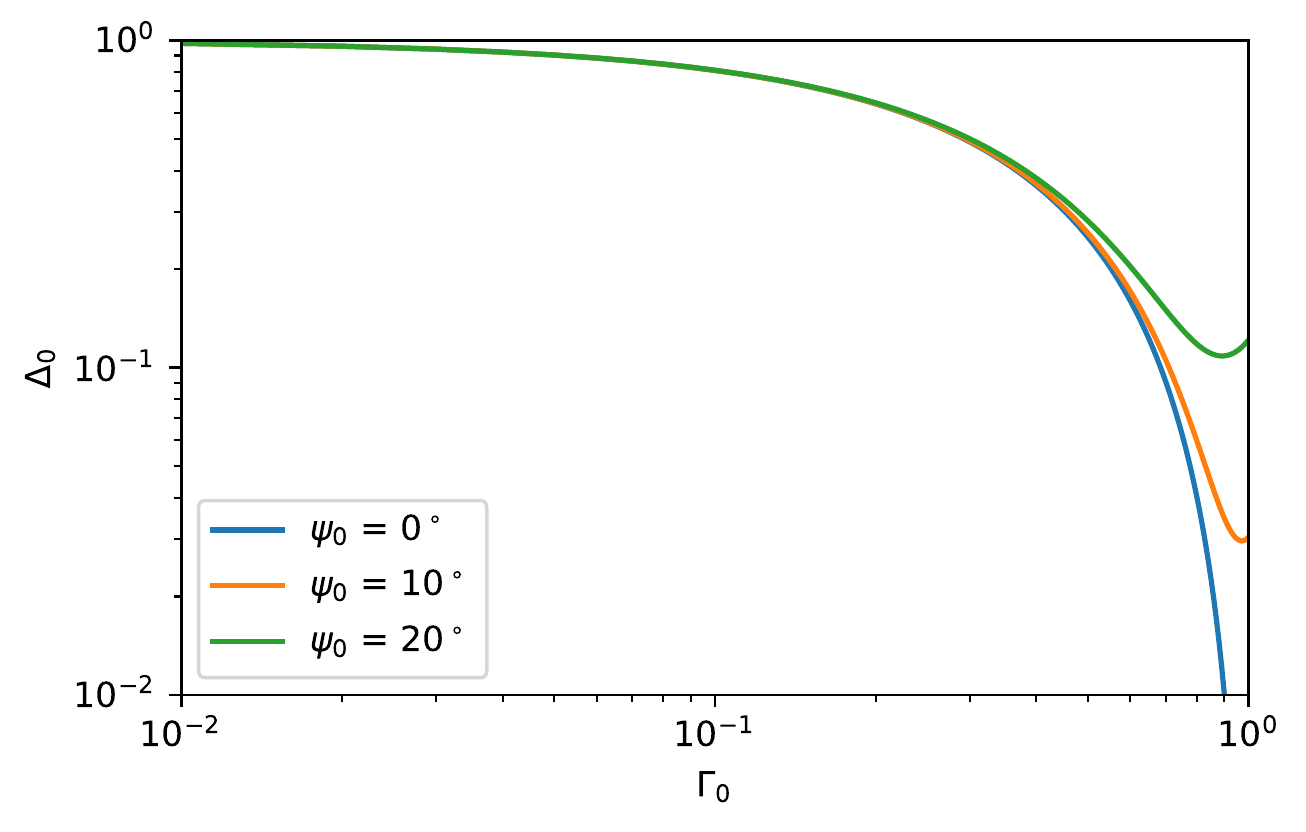} 
\caption{Value of $\Delta_0$ as a function of velocity amplitude reduction $\Gamma_0$ for various phase lags $\psi_0$.}
\label{fig:delta}
\end{figure}

Figure~\ref{fig:delta} shows $\Delta_0$ as a function of $\Gamma_0$ for various small phase lags $\psi_0$. It can easily be seen that if the motion is much smaller at the surface than in the water that $\Delta_0 \approx 1$ is a valid approximation. To first order, the phase lag is only important when there is negligible dissipation in the sea ice layer, which would occur in either very thin ice layers or more impermeable ice layers where $\epsilon h_i$ is very small. 
 
\subsection{Determining $\epsilon$}

As the determination of $\epsilon$ is most likely linked with the sea ice micro-structure this term is difficult to estimate. It seems reasonable that it would be related to the permeability of the sea ice, which is a function of ice temperature, salinity and ice volume fraction, and that using the law of fives~\citep{Golden_etal_2007}, which states that ice transitions from impermeable to permeable when the ice temperature reaches $-5^\circ C$ and bulk salinity is 5 parts per thousand, may provide a good starting point. For operational purposes it may be useful to estimate $\epsilon$ as a function of sea ice concentration, with lower sea ice concentrations having larger values of $\epsilon$, with the argument that more water will lead to more (relatively) warm sea ice which is more permeable. Such a parameterization would also be consistent with observations of wave attenuation being several orders of magnitude greater in the MIZ~\citep{Doble_etal_2015} than in pack ice~\citep{Wadhams_etal_1988,Meylan_etal_2014}. It can also be used as an empirical parameter to be fit a posteriori, not unlike current viscous ice models, from available data. 

For $\epsilon \to 1$, the rotational wave motion exists throughout the ice layer and the effective viscosity is given by $\nu = \omega h_i^2/2$. This scenario of $\epsilon \to 1$ provides an upper limit on the wave dissipation in the sea ice of $\alpha = h_i k^2/2$. In the limit as $\epsilon \to 0$, there is no wave dissipation within the sea ice and the ice layer is identical to that assumed by~\citet{Weber_1987}.

\section{Comparisons with previous experiments}
\label{sec:results}

Simultaneous observations of wave propagation and ice thickness are rare in the field and are predominantly restricted to laboratory experiments. In this section, (\ref{eq:alpha1}) is compared with available observations and it is assumed that $\Delta_0 = 1$, i.e. the ice motion is negligible compared to the water motion. However, without knowledge of the relative ice-water motion then the uncertainty in $\Delta_0$ is implicit in the uncertainty in the least-squares fit of $\epsilon$. 

Three laboratory experiments~\citep{Newyear_Martin_1997,Wang_Shen_2010_exp,Zhao_Shen_2015} and one field campaign~\citep{Doble_etal_2015} are used to validate (\ref{eq:alpha1}), as these experiments provide estimates of both the wave attenuation and ice thickness. The laboratory studies are performed with mechanically generated monochromatic linear waves, which allow for a direct comparison of our model with the observed wave dissipation. A summary of the least-squares fits with the laboratory data can be found in Table~\ref{tab:results}. Interpreting field data is decidedly more difficult as the wave field often consists of a broadband spectrum and the wind input and nonlinear terms can also have a non-negligible contribution to the spectral energy~\citep{Li_etal_2015,Li_etal_2017}. It is important to keep this in mind, that the wave dissipation due to the sea ice might not be the dominant process affecting the wave spectral energy, when comparing field observations with theoretical models for wave dissipation.

\begin{table}[h]
\centering
\begin{tabular}{l c c c}
\hline
   \textbf{Experiment} & \textbf{$\boldsymbol{h_i}$ / cm} & $\boldsymbol{\epsilon}$ & \textbf{R$\boldsymbol{^2}$} \\
\hline
\citet{Newyear_Martin_1997} & 11.3 & 0.70 $\pm$ 0.01 & 0.98 \\
                            & 14.6 & 0.60 $\pm$ 0.02 & 0.92 \\
\citet{Wang_Shen_2010_exp} & 9.0 & 0.56 $\pm$ 0.03 & 0.85 \\
                       & 8.9 & 0.64 $\pm$ 0.03 & 0.90 \\
\citet{Zhao_Shen_2015} & 2.5 & 0.34 $\pm$ 0.03 & 0.77 \\
                       & 4.0 & 0.94 $\pm$ 0.03 & 0.98 \\
                       & 7.0 & 0.85 $\pm$ 0.09 & 0.72 \\
\hline   
\end{tabular}
   \caption{Summary of fitting results using (\ref{eq:alpha1}) with data from available laboratory experiments. In all cases it is assumed that $\Delta_0 = 1$. \textbf{R$\boldsymbol{^2}$} denotes the coefficient of determination for the least-squares fit.}
   \label{tab:results}
\end{table}

\begin{figure}[t]
	\centering
   \includegraphics[angle=0]{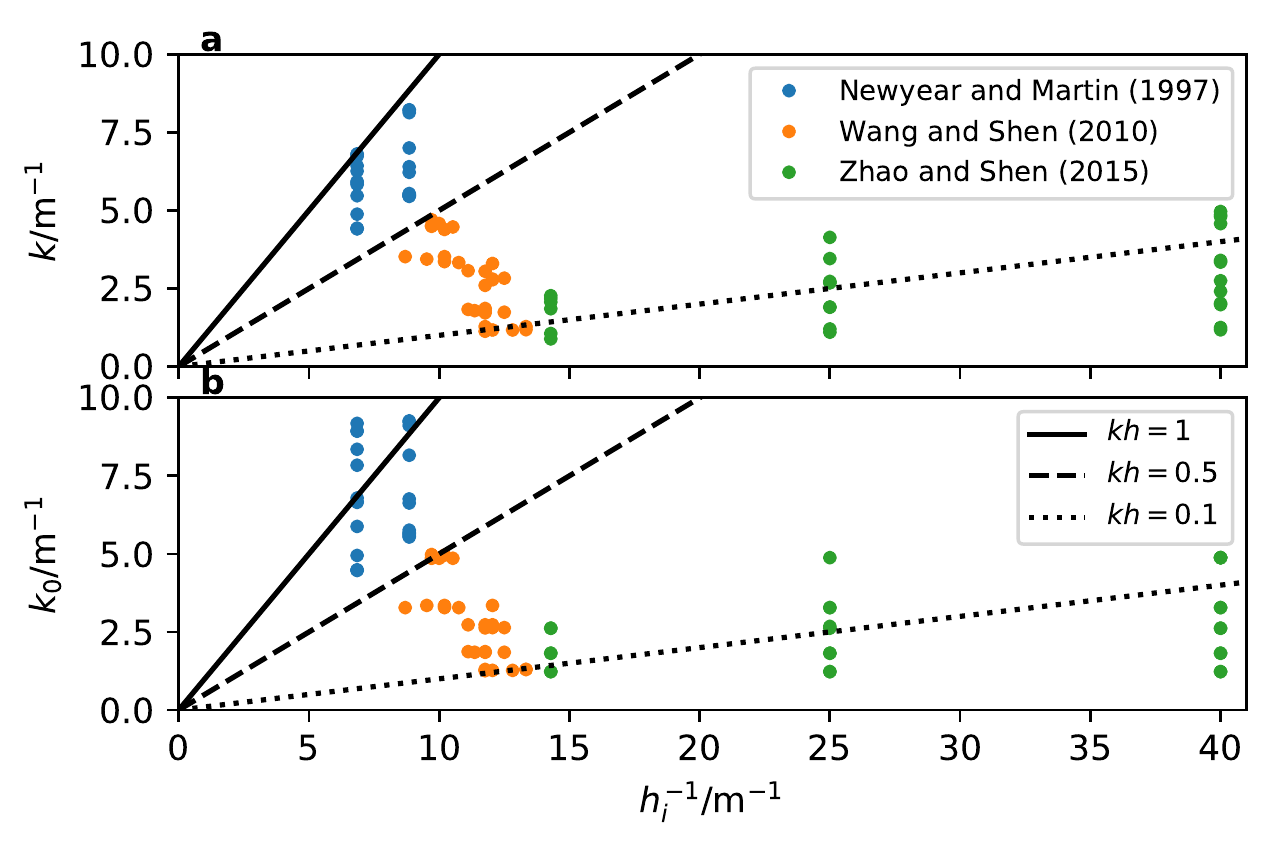} 
	\caption{\textbf{a} Observed values of the wavenumber $k$ as a function of inverse ice thickness $h_i^{-1}$ and \textbf{b} wavenumber calculated from the frequency using the open water relation $k_0 = \omega^2 / g$ as a function of inverse ice thickness. The black lines show various values for constant $k h_i$. The experiments correspond those in Table~\ref{tab:results}.}
	\label{fig:kh}
\end{figure}

It is often difficult to properly scale all the relevant wave-ice interaction processes in laboratory experiments~\citep[see][]{Rabault_etal_2018_jfm} given the geometric constraints of laboratory basins. Also, while field observations of the dispersion relation for waves in ice show little deviation from open water values~\citep{Marchenko_etal_2017}, this is not always the case in laboratory experiments~\citep{Newyear_Martin_1997}. Figure~\ref{fig:kh} shows $k$ vs $h_i^{-1}$ for the observed wavenumber (\ref{fig:kh}a) and the theoretical open water wavenumber $k_0 = \omega^2 / g$ (Figure~\ref{fig:kh}b).

Figure~\ref{fig:kh} shows relatively large values of  $kh_i$ in the~\citet{Newyear_Martin_1997} experiment with $k h_i \rightarrow 1$ for high frequencies with the effect being greater for the larger (smaller) ice thickness (inverse ice thickness). There is also a noticeable deviation from the open water value for these higher frequencies. In the experiment by~\citet{Wang_Shen_2010_exp}, $kh_i$ values are all smaller than in~\citet{Newyear_Martin_1997} with values ranging $0.1 < k h_i < 0.5$ and little deviation from the open water dispersion relation. The $k h_i$ values in~\citet{Zhao_Shen_2015} are smaller yet with these approximately equal to $k h_i =0.1$. 

\subsection{Laboratory experiment of \citet{Newyear_Martin_1997}}

\begin{figure}[t]
   \centering
   \includegraphics[angle=0]{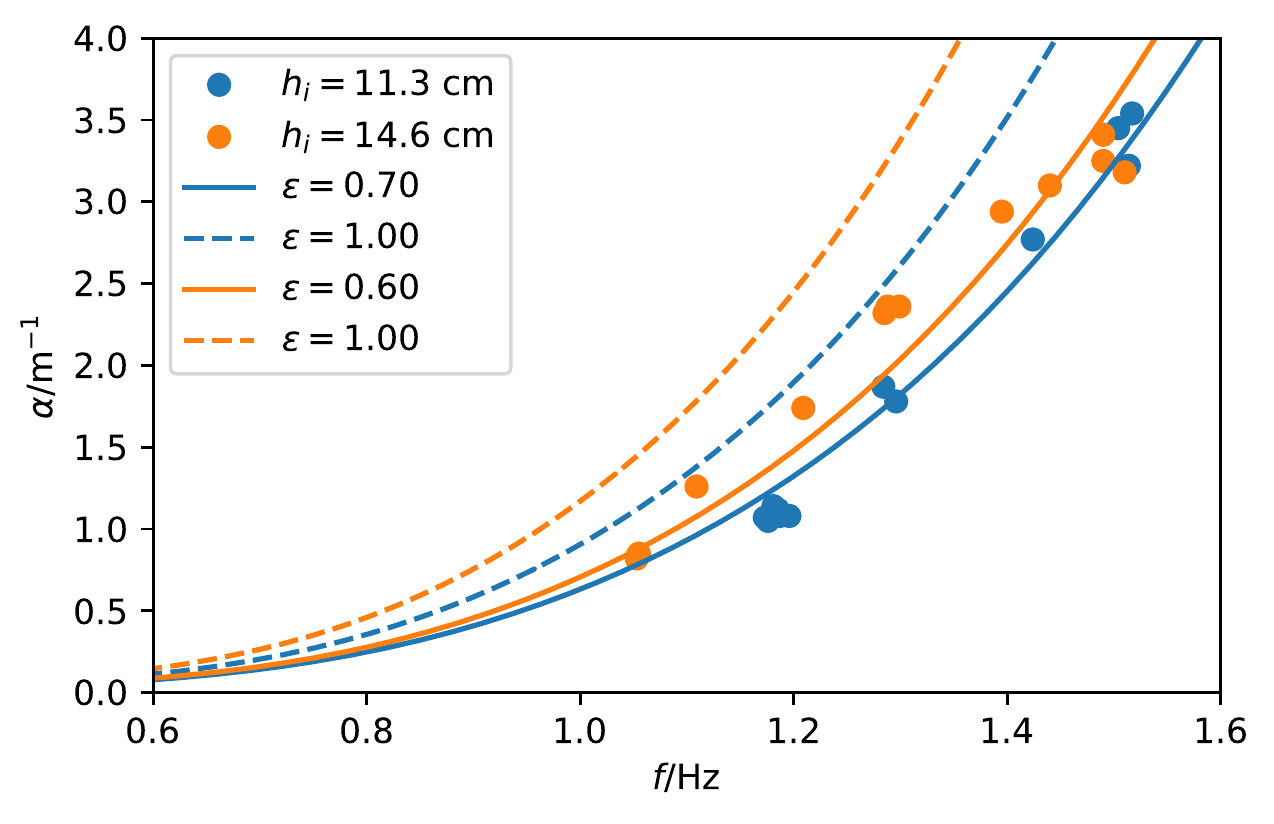} 
\caption{Comparison with results from~\citet{Newyear_Martin_1997}. Dots show observations for a given ice thickness $h_i$ and the curves show the results from (\ref{eq:alpha1}) for different values of $\epsilon$.}
\label{fig:nm97}
\end{figure}

This study was one of the first to measure wave dissipation due to frazil ice in a controlled laboratory experiment. Two experiments were performed with different frazil ice thicknesses:  one experiment was with an ice thickness of 11.3 cm and the other was 14.6 cm thick. Figure~\ref{fig:nm97} shows results for the wave dissipation from the two experiments, taken from Tables 1 and 2 in~\citet{Newyear_Martin_1997}. 

While the agreement between the observations and (\ref{eq:alpha1}) is qualitatively quite good, there is some deviation at the higher frequencies for the 14.6 cm experiment. At these high frequencies, $kh_i\approx 1$ (Figure~\ref{fig:kh}) and the assumptions leading to (\ref{eq:visc}), i.e that of a relatively thin layer with respect to the wavelength, are no longer be valid. These frequencies also correspond to where the dispersion relation deviated significantly from the open water relation for experiment 2 ($h_i$ = 14.6 cm) with the observed wavenumber being 70-80\% of the expected open water value~\citep{Newyear_Martin_1999}. For experiment 1 ($h_i$ = 11.3 cm) the maximum deviation of the dispersion relation relative to the open water is much less with the observed wavenumber being 90\% of the open water value at the highest frequencies.

\subsection{Laboratory experiment of \citet{Wang_Shen_2010_exp}}

This study was one of many performed as part of the REduced ice Cover in the ARctic Ocean (RECARO) project in the Arctic Environmental Test Basin at Hamburg Ship Model Basin (HSVA), Germany~\citep{Wilkinson_etal_2009}. Two experiments were performed with each having a very similar mean ice thickness of 8.9 cm and 9.0 cm respectively. 

\begin{figure}[t]
   \centering 
   \includegraphics[angle=0]{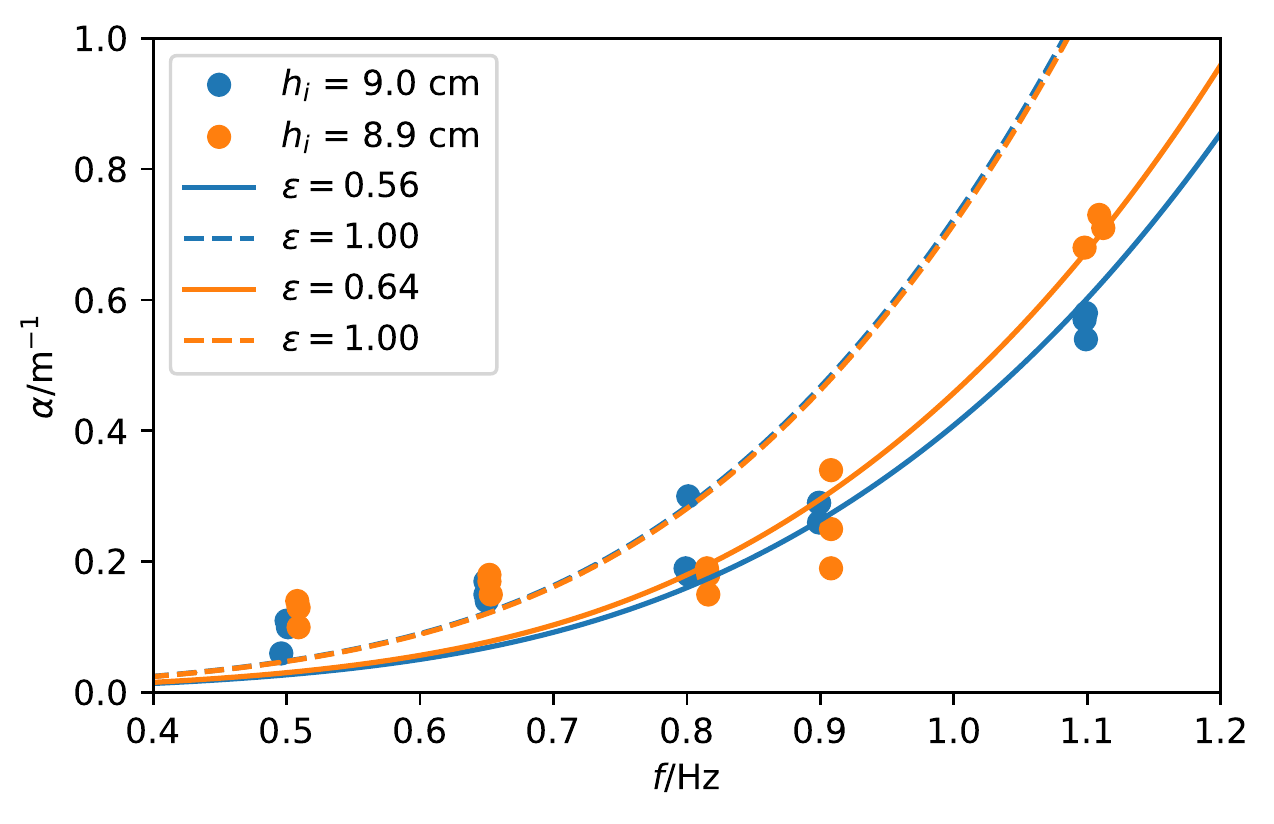} 
\caption{Comparison with results from~\citet{Wang_Shen_2010_exp}. Dots show observations for a given ice thickness $h_i$ and lines show (\ref{eq:alpha1}) for different values of $\epsilon$.}
\label{fig:ws10}
\end{figure}

The sea ice in this experiment is formed in a similar manner as~\citet{Newyear_Martin_1997}. A comparison of (\ref{eq:alpha1}) and the dissipation rates observed by~\citet{Wang_Shen_2010_exp}, from their Tables 1 and 2, can be found in Figure~\ref{fig:ws10}. Again, the comparison is quite good, but now the deviations from theory occur for the lower frequencies with the smallest dissipation rates. It is unclear why the dissipation is under-predicted at low frequencies, however the same deviation was also observed by~\citet{Wang_Shen_2010_exp} in fitting the viscoelastic model~\citep{Wang_Shen_2010}. The observed wavenumber in this experiment was similar to the open water value with a maximum deviation of 10\%.

\subsection{Laboratory experiment of \citet{Zhao_Shen_2015}}

\begin{figure}[t]
   \centering
   \includegraphics[angle=0]{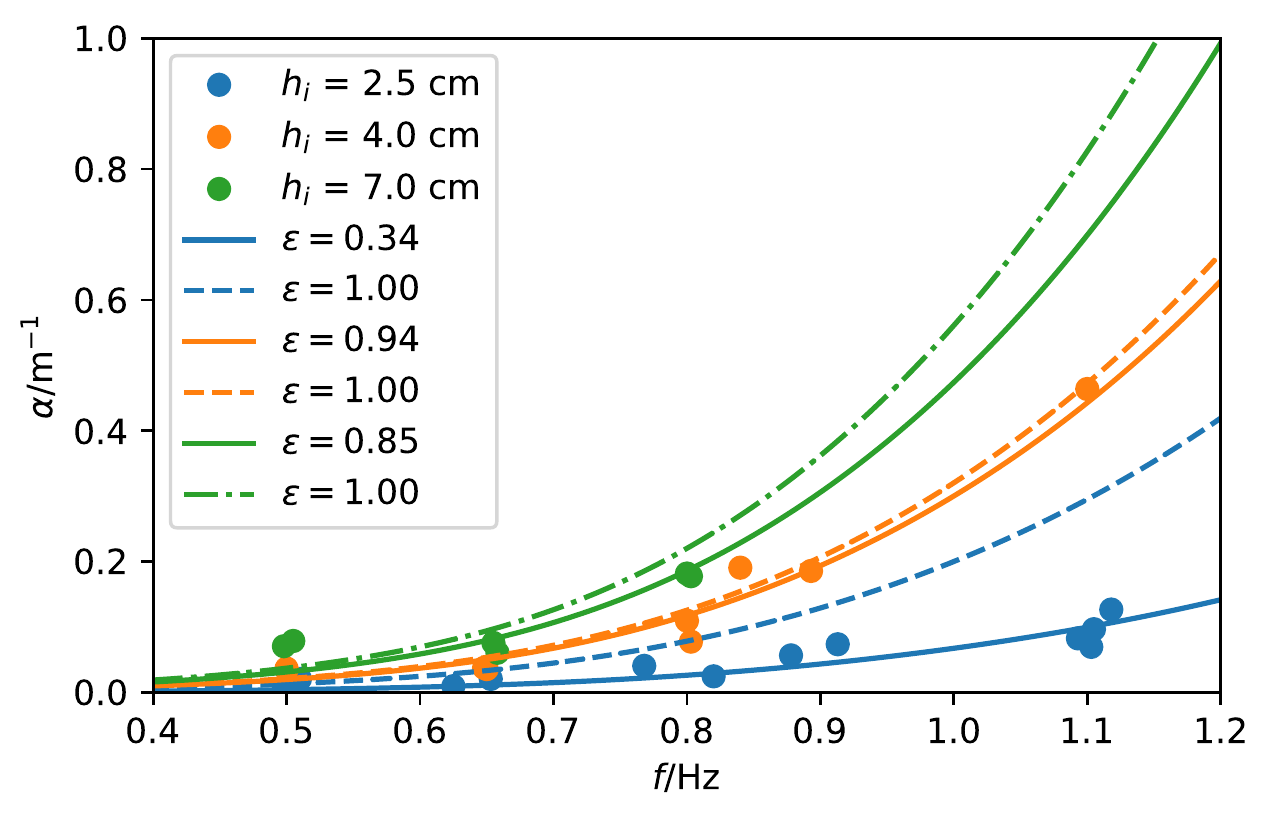}
   \caption{Comparison with results from~\citet{Zhao_Shen_2015}. Dots show observations for different ice thicknesses $h_i$ and lines show (\ref{eq:alpha1}) for different values of $\epsilon$.} 
\label{fig:zs15}
\end{figure}

These results were obtained from a subsequent experiment at the HSVA during 2013. In this experiment there were three distinct ice types: thin frazil ice (2.5 cm thickness), a frazil and pancake ice mixture (4.0 cm thickness) and a broken floe field (7.0 cm thickness). The results from each, as recorded in their Table 3, are shown in Figure~\ref{fig:zs15}.

Again, the comparisons are good with both the 4.0 cm and 7.0 cm thicknesses being well explained by (\ref{eq:alpha1}) for large values of $\epsilon$. For the thin frazil ice of 2.5 cm thickness, the observed wave dissipation had a very small $\epsilon$ of 0.34, which is most likely due to the relative motion of the ice, i.e. $\Delta_0 < 1$, which stands to reason would be greater for very thin ice layers than thicker layers. 

For the thicker ice, the same under-prediction of the wave dissipation relative to the observed value was observed, consistent with the results of~\citet{Wang_Shen_2010_exp}. \citet{Zhao_Shen_2015} also found that the viscoelastic model of~\citet{Wang_Shen_2010} under estimated $\alpha$ at low frequencies. 

\subsection{Field observations of \citet{Doble_etal_2015}}

Simultaneous field measurements of ice thickness and wave attenuation are rare, but a recent publication by~\citet{Doble_etal_2015} utilized detailed observations of the ice type and thickness~\citep{Doble_etal_2003} to investigate the wave attenuation characteristics for a frazil and pancake ice field in the Weddell Sea. \citet{Doble_etal_2015} found that the spatial energy attenuation rate was linearly related to the ice thickness according to
  \begin{equation}
    \label{eq:d15}
\alpha_*^{\mathrm{D}15} = 0.2 T^{-2.13}h_{\mathrm{eq}},
  \end{equation}
  where $T$ is the wave period and $h_{\mathrm{eq}}$ is an equivalent ice thickness, i.e. $h_{\mathrm{eq}}=\Phi h_i$ where $\Phi$ is the ice volume fraction. Note that the energy attenuation rate is related to the amplitude attenuation rate by $\alpha_* = 2\alpha$ and (\ref{eq:alpha1}) is multiplied by 2 for comparisons with~\citet{Doble_etal_2015}. The ice volume fraction used by~\citet{Doble_etal_2015} was 0.7 for pancake ice and 0.4 for frazil ice with estimates of the relative abundance of each in order to obtain a mean $h_{\mathrm{eq}}$. While (\ref{eq:alpha1}) predicts $\alpha\propto h_i$, it also predicts the dependence on the wave period to be $\alpha\propto T^{-4}$ in contrast with (\ref{eq:d15}), which has a period dependence of $\alpha\propto T^{-2.13}$. 
  
\begin{figure}[t]
   \centering
   \includegraphics[angle=0]{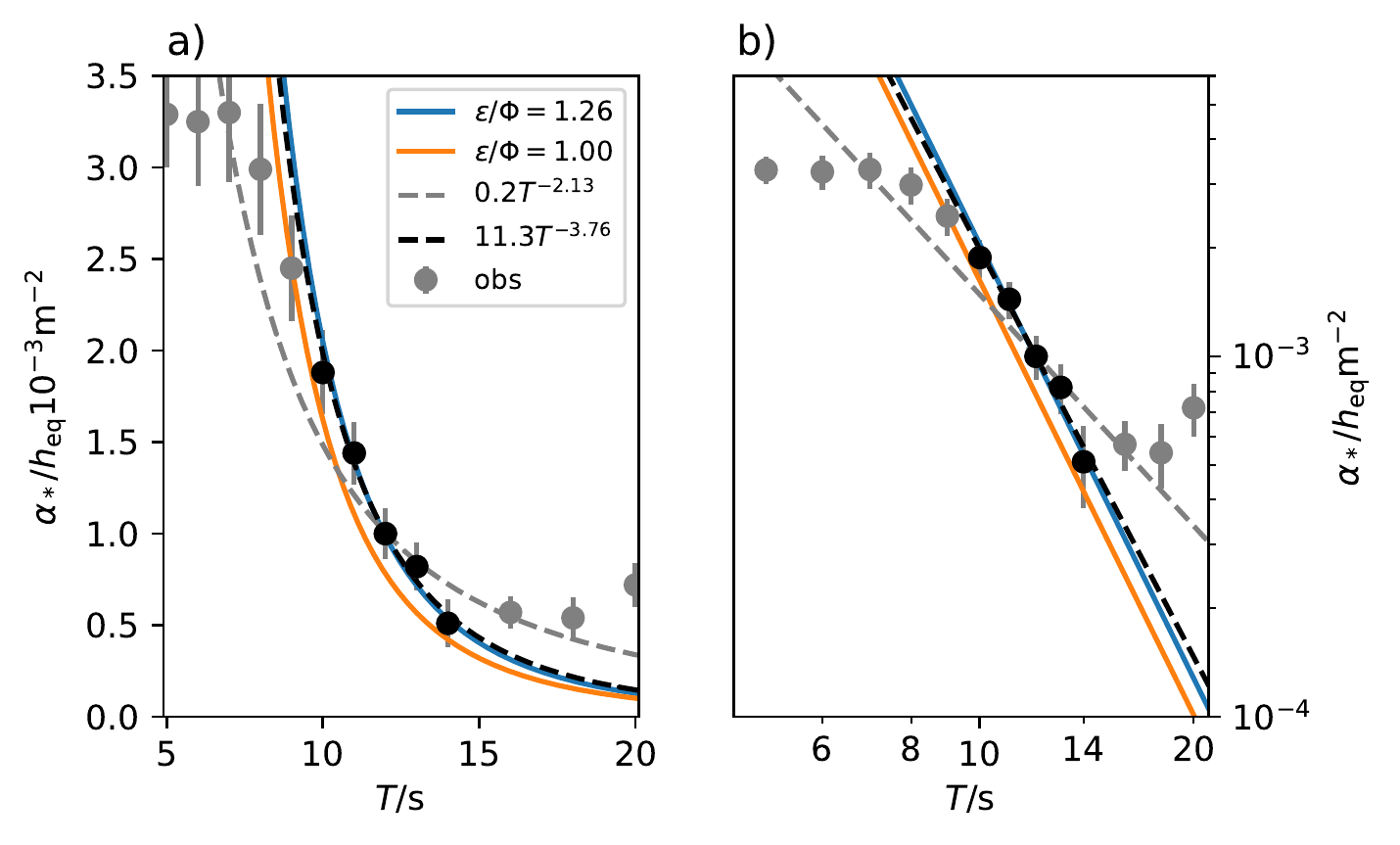}
\caption{Comparison with results from~\citet{Doble_etal_2015}. The equivalent ice thickness of (\ref{eq:d15}) is related with the ice layer thickness $h_i=\Phi^{-1}h_{\mathrm{eq}}$, where $\Phi$ is the ice volume fraction. a) linear scale plot and b) log-log plot showing the fit of \citet{Doble_etal_2015} (grey dashed line) and a power law fit (black dashed line) for a subset of frequencies (black dots). The y-axis shows the energy dissipation rate ($\alpha_*$), which is related to the amplitude attenuation rate by $\alpha_* = 2\alpha$.}
\label{fig:d15}
\end{figure}

  First, in order to compare (\ref{eq:d15}) with (\ref{eq:alpha1}), it is important to relate $h_{\mathrm{eq}}$ with the ice layer thickness $h_i$. This is important as (\ref{eq:alpha1}) depends on the thickness of the viscous layer, comprising a mixture of water and ice, and the thickness associated with only the ice component is not directly applicable. While the relative abundance of frazil and pancake ice is not known, we know it should be between 0.4 - corresponding to all frazil ice - and 0.7 - corresponding to all pancake ice. Using $h_{\mathrm{eq}}=\Phi h_i$ and (\ref{eq:alpha1}) multiplied by 2 to get $\alpha_*$ gives $\alpha_*/h_{\mathrm{eq}} = (\epsilon/\Phi) k^2$, which allows for a comparison of the observations of~\citet{Doble_etal_2015} and our model. 

  Second, it is clear from~\citet{Doble_etal_2015} that $\alpha_*/h_{\mathrm{eq}}$ is not a single power law and we must select a frequency range where wave dissipation is expected to be the dominant mechanism. It was shown by~\citet{Li_etal_2017} that the "rollover" in attenuation observed at low frequencies can be explained by the contributions from the wind input and the nonlinear transfer of energy and so attenuation is not expected to be the dominant mechanism. Therefore, we limit our analysis to wave periods above 9 s. In addition, \citet{Doble_etal_2015} state that "there is very little energy above 14 s, thus attenuation cannot be represented by these data", and thus we do not include the observations above 14 s. 
  
  Figure~\ref{fig:d15} shows the comparison of (\ref{eq:alpha1}), with different values of $(\epsilon/\Phi)$. Both a power law fit - black dashed line - as well as a least squares fit to (\ref{eq:alpha1}) - shown in blue - were performed and the corresponding coefficient of determination ($R^2$) values, which correspond to the fraction of variance represented by the least-squares fit, are 0.98 and 0.96 respectively. The power law fit gives a wave period dependence of $\alpha \propto T^{-3.76}$, with a standard error in the exponent of $-3.76 \pm 0.33$. This is consistent with (\ref{eq:alpha1}) which predicts $\alpha\propto T^{-4}$. The least-squares fit is for $\epsilon/\Phi$, which gives a best estimate of $\epsilon / \Phi = 1.26 \pm 0.05$. For pure grease ice $\Phi=0.4$ implies that $\epsilon=0.50$ and for an all pancake field $\Phi=0.7$ giving $\epsilon=0.88$. The true $\Phi$ is somewhere between these two limits, but the estimates of $\epsilon$ are consistent with laboratory experiments.

\section{Discussion}
\label{sec:discussion}

One uncertainty in using (\ref{eq:alpha1}) as a model for wave dissipation is the determination of $\epsilon$. However, since $\alpha$ in (\ref{eq:alpha1}) is linearly related to $\epsilon$, their relative uncertainties are also linear, i.e. a 50\% uncertainty in $\epsilon$ will correspond to a 50\% uncertainty in $\alpha$. Just to clarify, (\ref{eq:alpha1}) is a model for the wave dissipation in the ice layer and neglects viscous dissipation occurring below the ice - identical to the dissapitive models of~\citet{Keller_1998} and \citet{Wang_Shen_2010}. As $\epsilon\to 0$, the viscous dissipation below the ice will no longer be negligible compared to the dissipation occurring within the ice. This shortcoming can be addressed by adding the wave dissipation associated with the boundary layer under the ice~\citep[see][]{Liu_MolloChristensen_1988}, and this has been used in studies of wave dissipation for waves with long periods and small amplitudes over long distances into the pack ice~\citep{Ardhuin_etal_2016} where one would expect $\epsilon$ to be small as there will be negligible wave motion in solid ice. For example, \citet{Ardhuin_etal_2016} use the wave dissipation due to the boundary layer under the sea ice~\citep{Liu_MolloChristensen_1988}, which we will denote $\alpha^{\mathrm{LM88}}$ and is given by
\begin{equation}
   \alpha^{\mathrm{LM88}} = \frac{1}{2} d k^2
   \label{eq:alm88}
\end{equation}
where $d$ is the boundary layer under the ice calculated using $d = \sqrt{2\nu_w/\omega}$ and $\nu_w$ is the molecular kinematic viscosity of sea water. \citet{Ardhuin_etal_2016} multiply (\ref{eq:alm88}) by a factor of 12 in order to reproduce the observed wave heights. Since $d\propto \sqrt{\nu_w}$ a factor of 12 increase in $\alpha$ corresponds with a factor of 144 increase in $\nu_w$. However, it can easily be shown that $\alpha$ calculated using (\ref{eq:alpha1}) is related to $\alpha^{\mathrm{LM88}}$ by
\begin{equation}
   \alpha = \frac{\epsilon h_i}{d} \alpha^{\mathrm{LM88}},
   \label{eq:alpha_comp}
\end{equation}
Note that the frequency dependence of $d$ leads to $\alpha_{\mathrm{LM88}} \propto T^{-3.5}$, which is slightly different than (\ref{eq:alpha1}) where $\alpha\propto T^{-4}$. \citet{Ardhuin_etal_2016} used a molecular kinematic viscosity of sea water of $\nu_w = 1.83\times 10^{-6}$ m$^2$s$^{-1}$, which gives $d$ on the order of 10$^{-3}$ m for typical wave frequencies. Using an ice thickness on the order of 1 m gives an $\epsilon$ on the order of 10$^{-2}$ to reproduce the empirical factor of 12 used by~\citet{Ardhuin_etal_2016}. This shows that wave motion only needs to exist a few centimetres in the ice layer in order to produce much higher dissipation rates than would be expected in the boundary layer under the ice. This variability in $\epsilon$ may also explain the much greater wave dissipation observed in frazil and pancake ice~\citep{Newyear_Martin_1997,Doble_etal_2015,Rabault_etal_2017} compared to measurements in pack ice~\citep{Meylan_etal_2014,Ardhuin_etal_2016}.

The assumption of the no-slip condition may depend on the ice field as well as the size of the ice floes size as individual floes smaller than the wavelength may follow, or partially follow, the wave orbital motion\citep{Ardhuin_etal_2018,Marchenko_2018}. However, the dissipation of surface waves will apply a radiation stress~\citep{Longuet-Higgins_Stewart_1962,Weber_2001}, and this will act to compact the ice floes and limit their horizontal motion. The radiation stress is also expected to be responsible for the ice thickness gradient in laboratory experiments in grease ice~\citep{Martin_Kauffman_1981, Newyear_Martin_1997}. This radiation stress is proportional to the spatial attenuation rate so regions with high wave dissipation, such as the MIZ~\citep{Doble_etal_2015}, will have greater stresses compacting the floes. Direct observations of the relative water and ice floe motion in the MIZ are very rare, and the authors are only aware of one study by~\citet{Fox_Haskell_2001} which showed negligible horizontal motion of an ice floe relative to the vertical motion in the Antarctic MIZ. Furthermore, the wind stress, when aligned with the wave direction, will also cause a stress on the ice which would act to further compact the ice field and thus further limit the horizontal motion. However, the model can easily account for the relative motion of the ice and water with the $\Delta_0$ term in (\ref{eq:alpha1}). 

In comparing (\ref{eq:alpha1}) with the observations of~\citet{Doble_etal_2015} it was found that the wave dissipation dependence on wave period, $\alpha\propto T^{-4}$, was consistent with the observed dependence of $\alpha\propto T^{-3.76 \pm 0.33}$ when the period range was constrained to $10$ s $\le T\le14$ s in order to minimize the effects of other processes, such as wind input, on the on the spectral energy. This power law behavior is within the range of observed values for wave attenuation in sea ice~\citep{Meylan_etal_2018}. In addition, the magnitude over the selected frequency range also had a good agreement over suitable ice volume fractions $\Phi$ and $\epsilon$ for frazil and pancake ice. For wave dissipation, it appears that it is the geometry of the sea ice rather than the mass of the sea ice which is the important parameter. 

The dependence of the effective viscosity on ice thickness in (\ref{eq:visc}), i.e. $\nu\propto h_i^2$, is similar to that used in a recent study by~\citet{Li_etal_2017}, who employed an empirical formula for sea ice viscosity $\nu^{\mathrm{L17}} = 0.88h_i^2 - 0.015 h_i$ in a study on wave-ice interaction in the Southern ocean. In this quadratic equation for the ice viscosity, the linear term is much smaller than the quadratic term for ice thicknesses greater than 10 cm, thus the linear term is negligible in all but the thinnest of ice covers. In addition, the quadratic equation predicts a negative viscosity for $h_i<1.7$ cm, which is clearly not physical. In calm conditions, where the significant wave height was less than 1 m, \citet{Li_etal_2017} found that mean errors were reduced by 30\% using $\alpha^{\mathrm{L17}}$ compared to a constant viscosity. However, it should be noted that the viscoelastic model of~\citet{Wang_Shen_2010} was employed and it is unclear as to the extent the elasticity is contributing to the results.

It is also of interest to note that the wavelength dependence of (\ref{eq:alpha1}) is identical to that derived by~\citet{Kohout_etal_2011}, which was estimated from the bottom drag due to the roughness of ice floes. Strictly speaking, the bottom roughness does not predict an exponential decay, so their spatial amplitude dissipation rate was calculated assuming that an exponential, or approximately exponential, dissipation rate could be assumed over a relatively short distance. This assumption gives an estimated dissipation rate of
  \begin{equation}
    \label{eq:k11}
    \alpha^{\mathrm{K11}} = 2H_SC_dk^2,
  \end{equation}
  where $H_S$ is the significant wave height and $C_d$ is the drag coefficient, which can be tuned to observations and is a function of the underside roughness. In a study in the Antarctic, \citet{Doble_etal_2013} used (\ref{eq:k11}) to calculate the wave dissipation by sea ice, along with the scattering model of~\citet{Kohout_Meylan_2008}, with reasonably good results. However, it does appear that the modelled attenuation is overestimated, relative to buoy observations, for large $H_S$ and underestimated for small $H_S$. \citet{Doble_etal_2013} used a value of $C_d=10^{-2}$, which for their observed significant wave heights between 10 m and 20 m, provides an estimate of sea ice thickness from (\ref{eq:alpha1}) between 20 cm and 40 cm (assuming $\epsilon=\Delta_0=1$), and is remarkably consistent with their estimate of ice thickness of $h_i = 0.2 + 0.4C$ where $h_i$ is the ice thickness in metres and $C$ is the sea ice concentration. While their relation between $C$ and $h_i$ is clearly not valid for $C=0$, as the parameterization would yield $h_i=0.2$ m where there should be no ice, it does reproduce the thickness of the 30\% ice concentration ($h_i = 0.32$ m) with good accuracy. This 30\% contour thickness would give similar estimates for the attenuation between (\ref{eq:alpha1}) and (\ref{eq:k11}), suggesting that their good agreement could be due to their particular ice and wave conditions. Further experiments would be required to test the validity of (\ref{eq:k11}) in seas with smaller amplitude waves as well.

While comparisons between observations and (\ref{eq:alpha1}) are generally quite good, there still exist outstanding questions with regards to the applicability to larger floes, as the flexure of these floes create other dissipative mechanisms~\citep{Marchenko_Cole_2017}, in addition to having a different dispersion relation which acts to effect the group velocity relative to that of open water~\citep{Sutherland_Rabault_2016}. One of the mechanisms that~\citet{Marchenko_Cole_2017} investigate is wave dissipation due to brine migration that is created by the flexure of large flows, and that this could potentially be a large sink of wave energy. Nonlinear processes could also be a contributing factor due to the different dispersion relation creating conditions for wave-triad resonance~\citep{Deike_etal_2017}, which were shown to be more efficient than the four-wave resonance of water waves. Another contributor to wave dissipation could be the relative motion of the ice floe with respect to the water could lead to large turbulent eddy viscosities below the sea ice~\citep{Marchenko_etal_2017,Marchenko_Cole_2017}, which may increase the dissipation of wave energy~\citep{Liu_MolloChristensen_1988}.

\section{Conclusions}
\label{sec:conclusions}

Presented is a new method to estimate wave dissipation under sea ice. We have derived an "effective viscosity" using dimensional analysis where the viscosity should scale as $\nu\propto h_i^2$. This scaling leads to an amplitude attenuation rate scaling of $\alpha\propto k^2$. Comparisons are made with available laboratory and field experiments that measured both ice thickness and wave dissipation. The model only accounts for wave dissipation within the ice layer, identical to the model of~\citet{Wang_Shen_2010}, and numerical models would need to include other sources of dissipation such as scattering due to individual ice floes~\citep{Kohout_Meylan_2008} and friction at the base of the ice layer~\citep{Ardhuin_etal_2016}. The strong agreement of our model with available laboratory data, where other sources of dissipation are expected to be negligible, is encouraging.

While the model predicts the linear dependence of attenuation on ice thickness, as suggested by~\citet{Doble_etal_2015}, and predicts a power-law dependence of $k^2$, consistent with observed power laws~\citet{Meylan_etal_2018}, there is still the question of determining $\epsilon$ and $\Delta_0$ in order to predict the magnitude of the attenuation. For most cases it appears that $\Delta_0 \approx 1$ is a good approximation, although an empirical function of wavelength and floe size, similar to~\citet{Ardhuin_etal_2018}, could also be implemented. This leaves the determination of $\epsilon$, which would naturally seem to be related to the ice micro-structure~\citep{Golden_etal_2007}, although further studies would be required to investigate this. Both $\Delta_0$ and $\epsilon$ have a maximum value of 1, which does allow for an estimation of the upper limit of wave attenuation. Parameterizations for $\Delta_0$ and $\epsilon$ based on the ice field (e.g. ice concentration, floe size distribution) and/or the wave field could be explored for modelling wave propagation under sea ice. The model is also simple to implement and avoids many of the difficulties associated with more complex models for wave propagation in sea ice~\citep{Mosig_etal_2015}. 

%
\section{Acknowledgements}
All data used in the study are available from the cited references. This project is funded by the Norwegian Research Council (233901) and through CMEMS ARC-MFC.

%

%
\bibliographystyle{jfm}

\end{document}